\documentclass[galaxies,article,accept,pdftex,moreauthors]{Definitions/mdpi} 

\firstpage{1} 
\pubvolume{1}
\issuenum{1}
\articlenumber{0}
\pubyear{2023}
\copyrightyear{2023}
\datereceived{30 April 2023} 
\daterevised{6 June 2023} 
\dateaccepted{7 June 2023} 
\datepublished{ } 
\hreflink{https://doi.org/} 

\Title{On the Evolution of, and Hot Gas in, Wind-Blown Bubbles around Massive Stars -
Wind Bubbles Are Not~Energy-Conserving} 
\TitleCitation{On the Evolution of, and Hot Gas in, Wind-Blown Bubbles around Massive Stars: Wind Bubbles Are Not Energy-Conserving} 

\Author{Vikram V. Dwarkadas
 \orcidA{}}

\AuthorNames{Vikram V. Dwarkadas} 

\AuthorCitation{Dwarkadas, V. V.
}

\address[1]{%
Department of Astronomy and Astrophysics, University of Chicago, Chicago, IL 60637, USA; 
 vikram@astro.uchicago.edu}

\abstract{The structure and evolution of wind-blown bubbles (WBBs) around massive stars has primarily
 been investigated using an energy-conserving model of wind-blown bubbles. While this model is useful in explaining the general properties of the evolution, several problems remain, including inconsistencies between observed wind luminosities and those derived using this formulation. Major difficulties include the low X-ray temperature and X-ray luminosity, compared to the model. In this paper, we re-examine the evolution, dynamics, and kinematics of WBBs  around massive stars, using published ionization gasdynamic simulations  of wind-blown bubbles. We show that WBBs can cool efficiently due to the presence of various instabilities and turbulence within the bubble. The expansion of WBBs is more consistent with a momentum-conserving solution, rather than an energy-conserving solution. This compares well with the dynamics and kinematics of observed wind bubbles. Despite the cooling of the bubble, the shocked wind temperature is not reduced to the observed values. We argue that the X-ray emission 
arise mainly from clumps and filaments within the hot shocked wind region, with temperatures just above 10$^6$ K. The remainder of the plasma can contribute to a lesser extent.}


\newcommand{\beqn}{\begin{eqnarray}}
\newcommand{\eeqn}{\end{eqnarray}}
\newcommand{\be}{\begin{equation}}
\newcommand{\ee}{\end{equation}}

\newcommand{\msun}{\mbox{$M_{\odot}$}}

\newcommand{\bfig}{\begin{figure}[H]}
\newcommand{\efig}{\end{figure}}

\newcommand{\apss}{\mbox{ApSS}}

\begin{document}

\section{Introduction}

Massive stars ($\geq$10 $\msun$) lose mass throughout their lifetime via
stellar winds and outbursts. They will either end their lives  in a cataclysmic
supernova (SN) explosion or~collapse directly to a black hole in the event of a failed SN~\citep{sukhboldetal16}.~(
{However, see \citep{soker17} for an alternate scenario}) The~interaction of
 the expelled material with the
surrounding medium creates vast wind-blown cavities surrounded by a
dense shell, referred to as wind-blown bubbles (WBBs), which are ionized by the hot UV photons from the star. As~the star
evolves through various stages, the~mass-loss parameters, and~the number of ionizing photons, will change. This affects the structure and evolution of the bubble. If~the star explodes as a SN, the~resulting SN shock wave will expand within the bubble, and~the dynamics and kinematics of the shock wave will depend on the bubble parameters \citep{cs89,tenorioetal1990,tenorioetal1991,dwarkadas05, dwarkadas07}. Similarly, the~relativistic blast waves associated with gamma-ray bursts (GRBs) should expand within wind bubbles surrounding Wolf--Rayet (W-R) stars \citep{wb06}. Thus, the~evolution and emission from SNe and GRBs are affected and influenced by the presence of WBBs, making them important to~understand.

Modeling of nebulae around massive stars has been ongoing 
 since at least the work of \citet{avedisova72}. The~structure of WBBs was clearly outlined in the seminal work of \citet{weaveretal77}. Proceeding outward in radius from the star, they identified four different regions: (1) A freely expanding wind region. If~the wind velocity  $v_w$ and the mass-loss rate  $\dot{M}$ are constant, then the density in this region ${\rho}_w = \dot{M}/(4 \pi v_w r^2)$ decreases with radius $r$ as $r^{-2}$. (2) A shocked wind region, separated from the freely expanding wind by a wind-termination shock. (3) The shocked ambient region, which forms a thin, dense shell. The~inner boundary of this region is a contact discontinuity, separating the shocked wind and shocked ambient medium. The~outer boundary is a shock, which for any reasonable wind parameters is a radiative shock. 
  (4) The unshocked ambient medium, which could be another wind or the interstellar medium. Most of the volume of the bubble can be shown to be occupied by the high pressure, low density shocked~wind.

In order to accurately understand the structure of WBBs around massive stars, models must take into account both the gasdynamics and photoionization due to stellar photons. Analytic theory of wind bubbles around massive stars was explored in detail by \citet{km92a,km92b}. \citet{chevalier97} studied the expansion of a photoionized stellar wind, although~mainly in the context of planetary nebulae. Early models that explored the evolution of massive star surroundings \citep{glm96, gml96} had some ionization built in, primarily centered around the Strömgren sphere approximation. They did not model the main-sequence (MS) stage in multi-dimensions, and~they did not treat recombination accurately, if~at all. Modeling of the Homunculus Nebula  around $\eta$ Car was carried out by \citep{fbd95,db98,frd98}. Bipolar wind bubbles around Luminous Blue Variable 
nebulae, arising from radiatively driven winds, were explored in \citep{do02}. Models of \citep{vanmarleetal05, vanmarleetal06} included limited treatment of ionization. A~somewhat better treatment of ionization was included in \citep{fhy03, fhy06}.  
 Dwarkadas~\citep{dwarkadas07a, dwarkadas07, dwarkadas08} considered the entire evolution of wind-blown nebulae in multi-dimensions, including the MS phase, and studied the turbulence in the interior, but did not include any photoionization in the calculations.
 3D simulations carried out by \citep{vanmarleetal11} also did not include the effects of ionizing photons. The~first works that included a reasonably accurate treatment of ionization
 from stellar photons, as well as gasdynamics, were carried out by Arthur et al. \citep{arthur07, arthur09}. A~comprehensive paper showing the evolution of wind bubbles around massive stars, including both photoionization and gas dynamics, was by \citep{ta11}.  Ionization-gasdynamic simulations were also carried out by \citep{dr13}, which were further investigated in \citep{vvd22}. The~relative impact of photoionizing radiation versus stellar winds was studied by \citep{ck01,haidetal18}. Stellar wind bubbles in an HII region were studied by \citep{geenetal21}, while those around a cluster of stars were modeled by \citep{lancasteretal21a,lancasteretal21b}. Bubbles around W-R stars have been modeled, without~including photoionization, by~\citep{meyeretal20, meyer21}.

One-dimensional simulations are in good agreement with the dynamical and kinematical picture 
 presented in the model by \citet{weaveretal77}. Multi-dimensional simulations are subject to instabilities, turbulence, and departures from symmetry, thus adding more complexity, while remaining qualitatively true to the basic picture. The~wind bubble theory has, in general, been quite successful at outlining the evolution and kinematics of wind bubbles around massive stars,  and~even superbubbles around clusters of stars \citep{mm88}. However, over~the years, it has become clear that there also exist major disagreements between the observations and the predictions of the model.

\begin{itemize}
\item As pointed out by authors such as \citet{nazeetal01}, some observed bubbles do not appear to conform to the theoretical predictions. Instead, wind mechanical luminosities ($L = 0.5 \dot{M} \;v_w^2$), calculated using the \citet{weaveretal77} model, can be up to two orders of magnitude lower than those expected from the stellar parameters. 
\item Wind bubbles should technically be found around every massive star since they all have fast winds. However, they are rarely seen around main-sequence O and B stars~\citep{chu08}. 
\item \textls[-25]{The wind velocities of  O, B, and W-R
stars are of the order of \mbox{1000--3000 km s$^{-1}$}.} The~post-shock temperature $T_w$ in the shocked wind region would then be
\begin{equation}
T_w=\frac{3}{16} \frac{\mu m_H}{k_b}\,v_w^2 \approx 1.36 - 12.25 \times 10^7 K. 
\end{equation}

\noindent
where  $\mu$ is the mean molecular weight of the gas,  $k_b$ the Boltzmann constant, and $m_H$ is the mass of a hydrogen atom. $\mu \approx 0.6$ for a fully ionized gas has been used. In~this calculation, the~wind termination shock is assumed to be slowly moving in the laboratory frame, which is generally~true.

Given the expectation of such high temperatures in this large volume of shocked wind, wind-blown nebulae should be visible as regions of diffuse X-ray
emission,
with~X-ray temperatures on the order of 1--10 keV. However, although~extensive searches have been carried out using Chandra and XMM,
diffuse X-ray emission has 
 been detected in only a few cases \citep{chu03,chuetal03,cgg03,wriggeetal05}. Even in those cases, the observed X-ray
temperatures are \mbox{10--100 times} smaller than what would be expected from the \citet{weaveretal77} model. For~the WBBs NGC 6888 and S308 surrounding W-R stars, the~inferred X-ray  temperatures are a few times 10$^6$ K, which is lower than expected. In~the case of NGC 6888,  a~higher temperature
component (>2 keV) inferred from {\it Suzaku} data by \citet{zp11} is not supported  by  {\it Chandra} and {\it XMM-Newton} data \citep{toalaetal14, toalaetal16}. A~high temperature component (>4.5 keV) for NGC 2359 was found by \citet{toalaetal15}, but~the contribution of this component to the total X-ray flux was less than 10\%.  Diffuse X-ray emission has also been seen in the W-R nebula NGC 3199 around the W-R star WR 18~\cite{toalaetal17}. Here, again, the dominant plasma temperature is around 0.15 keV, with~a hotter component contributing less than 8\% of the flux. Deep X-ray observations of NGC 7635 \citep{toalaetal20} failed to reveal any signs of X-ray emission. Many other wind-blown bubbles around massive stars do not show the presence of X-ray emission at all \citep{chuetal06,chu08}. 

\end{itemize}

The lack of observable diffuse X-ray emission, and~the low temperatures when X-ray emission is detected, could suggest the existence of a mechanism that  lowers the interior temperature or,~equivalently, the energy that goes into raising the temperature is being expended elsewhere. Or,~it could indicate that our expectations and assumptions are~flawed.

In this paper, we explore various facets of the evolution of wind-blown bubbles around massive stars to understand their dynamics and kinematics, as well as~the X-ray emission, using results derived from ionization gas dynamic simulations of wind-bubbles around massive stars, published in \citep{vvd22}. In~Section\ref{sec:sim}, we summarize the results from prior  simulations of wind bubbles.  In Section~\ref{sec:rad}, we study the bubble expansion, and~show that it does not fit the pressure-driven solution proposed by \citet{weaveretal77}. A momentum-conserving solution provides a better fit. Section~\ref{sec4} investigates various factors that would cause the loss of the energy reservoir and affect the bubble dynamics and kinematics. The~implications of this on prior results related to wind-blown bubbles are discussed in Section~\ref{sec5}. Section~\ref{sec:Xray} investigates the X-ray emission from wind bubbles in the context of our results. Conclusions are detailed in Section~\ref{sec7}. 

\section{Results from Ionization-Gasdynamic Simulations}
\label{sec:sim}

\begin{figure}[H]
\includegraphics[width=0.98\columnwidth]{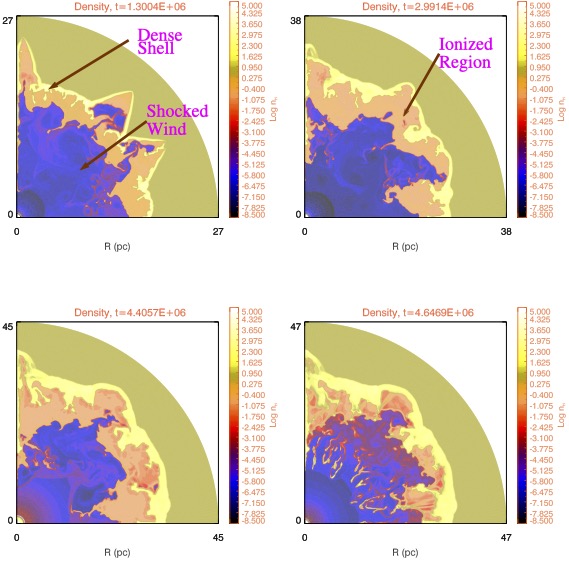}
\caption{Snapshots of the number density (in cm$^{-3}$) with time. From the calculation of the evolution of a wind bubble
  around a 40 $\msun$ star (600 $\times$ 400 zones), computed using an expanding grid (\cite{dr13, vvd22).}
  Time increases from left
  to right and top to bottom. The time in years is listed at the top of each panel. The scale shows the log of the number density. The~dense swept-up shell, the~ionized HII region, and~the shocked wind region are all marked in the figure. The~top two panels depict the MS phase. In~the bottom left one, the RSG wind can be seen expanding near the star. It does not go too far given its low velocity. The~bottom right panel depicts the clumps and filaments in the shocked wind region during the W-R phase.
\label{fig:hydro}}
\end{figure}

An exploration of the 2D hydrodynamics around a 40 $\msun$ star, including a proper treatment of the ionization and recombination along with the hydrodynamics, was carried out by \citet{dr13}. \citet{vvd22}~(hereafter, Paper 1) explored the ionization properties of the bubble, and~the differences between 1D and 2D models, in~further detail. The~star in these simulations evolves from the MS to the red supergiant (RSG) phase, and~ends its life as a W-R star.  The~entire evolution, including the MS phase, was carried out in 2D, unlike most simulations that do not simulate the MS phase in 2D. In~order to aid in the subsequent discussion, we summarize the evolution of the bubble around a massive star. Full details can be found in Paper 1.  Density snapshots from the evolution are shown in Figure~\ref{fig:hydro}.
 
The simulations were carried out using the ionization gasdynamics code AVATAR. The code operator-splits the contribution due to photoionization from the gas dynamics, and~utilizes a backward-Euler scheme together with a Newton--Raphson iteration procedure. The effects of geometrical dilution and column absorption of radiation are taken into consideration. A~second-order monotonic transport algorithm,  or~a third-order piecewise parabolic scheme, is used for the advection of the total mass and neutral component. Tabulated functions are used to compute the collisional ionization rate and cooling function. Shocks are treated using artificial viscosity. The option of grid expansion is available. The~algorithm incorporates a model of the photoionization source, computes the fractional ionization due to the photoionizing flux and recombination, and~self-consistently determines the energy balance by considering ionization, photo-heating, and radiative~cooling.

In this paper, we further investigate the 2D simulation described in Paper 1. This simulation computes the evolution of the wind-blown nebula around a 40 $\msun$ star,  including stellar photoionization as well as radiative cooling, using the AVATAR code.  It was run in spherical co-ordinates on a grid with 600 radial and 400 angular zones. The~grid is not adaptive, but~it is expanding---as the shock expands outwards, the~grid expands with it, so the grid size increases over time, as seen in the figures. This allows for adequate grid resolution throughout compared to a fixed grid. All analyses and figures in this paper refer to this simulation. The stellar parameters used to model the wind-blown nebula are adapted from~\cite{vanmarleetal05}. The~evolution of the star can be divided into three main phases. The~star starts its life as an O star in the MS phase. An~inhomogeneous pressure and density distribution develop, accompanied by vorticity deposition near the inner shock. The~inclusion of photoionization results in the formation of a dense, lower temperature ($\sim$10$^4$ K) region of ionized material outside the hot shocked wind region during the MS phase. The~nebula is fully ionized, and~the ionization front is found to be unstable to various instabilities. The~hot shocked wind region slowly begins to assume an aspherical geometry. In~the RSG stage, the surface temperature of the star decreases considerably. Consequently, the ionizing radiation from the star drops significantly, and~recombination reduces the ionization fraction in the ionized HII region to $\sim$30\%, although~this rises again in the W-R stage. The~high-density RSG wind is followed by a higher momentum W-R wind. This tends to break up the RSG wind material, forming clumps and filaments that are mixed into the hot shocked wind. The~geometry of the hot shocked wind region remains aspherical  throughout.

In simulations that include a treatment of the ionization properties, as  in~\cite{ta11,dr13,vvd22}, the structure of the bubble is found to deviate from the model of \mbox{\citet{weaveretal77}}. A~higher density ionized region, formed by the photons from the star ionizing the surrounding material, in~this case, the interstellar medium, is seen. The~size of any such region would depend on the star's surface temperature and the number of ionizing photons from the star.  Although~
 these would be difficult to observe, an~ionized HII region inner to the equatorial ring was postulated in the structure of the wind bubble surrounding SN 1987A, in order to explain the increasing X-ray and radio emission from the SN ~\cite{cd95}. The~ionized region ends in a contact discontinuity, outside of which lies the dense shell of  swept-up surrounding medium, bounded on the outside by a radiative shock. The~ionized region is found to be unstable to various instabilities, including ionization front instabilities, combined with Rayleigh--Taylor and Kelvin--Helmholtz instabilities at the inner edge of the dense shell and in the shocked wind region (Paper 1).

In this paper, when we refer to the `bubble', we mean the entire structure, i.e., the hot shocked wind, the ionized region, and the dense shell. In~practice, this is what will be seen by an observer and referred to as the wind bubble. When referring 
 to the radius of the bubble, it will be the outer radius of this~structure.

\section{Bubble Radius and Expansion}  
\label{sec:rad}

The \citet{weaveretal77} solution predicts that the bubble  radius $R_b$ increases as
\begin{equation}
R_b = \xi \left[\frac{L}{\rho_{a}}\right]^{1/5}\;t^{3/5}
\end{equation}

\noindent
where $L = 0.5 \dot{M} v_w^2 $ is sometimes referred to as the mechanical luminosity of the wind,  $\rho_{a}$ is the density of the ambient medium, and~$t$ is the age of the bubble. The~
 coefficient $\xi$ was found by \citet{weaveretal77} to be $(250/308\pi)^{1/5} = 0.76$. We refer to this as the pressure-driven solution, as~in~\cite{lancasteretal21a}.

As explained below, energy can be radiated away or expended in turbulent motions, reducing the energy reservoir available for the expansion of the bubble. The~radius of the bubble  can consequently become smaller than that given by the \citet{weaveretal77} solution.

\citet{elbadryetal19} studied the evolution of bubbles, where there was some amount of cooling. They assumed that a fraction $\theta$ of the wind luminosity was lost to radiation due to turbulent mixing between the dense shell and hot interior. Their solution straightforwardly modifies the \citet{weaveretal77} solution, giving
\begin{equation}
R_b = \xi \left[\frac{L(1-\theta)}{\rho_a}\right]^{1/5}\;t^{3/5}
\end{equation}

The bubble still expands with the same time dependence, albeit with a lower mechanical luminosity;~thus, the radius will be smaller than in the pressure-driven case.

If cooling is so efficient that the pressure-driven solution is no longer relevant, a~momentum-driven phase will be reached. The shocked wind impacts directly on the dense shell or photoionized region.  The~radius of a momentum-conserving (MC) bubble can be shown to increase with time as \citep{km92a,km92b,lc99}:
\begin{equation}
R_{MC} = \left[\frac{3}{2\pi}\frac{\dot{M}\,v_w}{\rho_a}\right]^{1/4}\;t^{0.5} 
\label{eq:ec}
\end{equation}

In the momentum-driven case, the bubble radius grows as $R_b \propto t^{0.5}$.

\citet{lancasteretal21a} modified this solution to accommodate a  bubble with a higher momentum ${\alpha}_p \dot{p}$ compared to the initial input momentum $\dot{p}$, and~an order unity parameter ${\alpha}_R$, to~give
\begin{equation}
R_{b} = ({\alpha}_R \, {\alpha}_p)^{1/4} \; R_{MC}
\end{equation}

\begin{figure}[H]
\includegraphics[width=0.62\columnwidth]{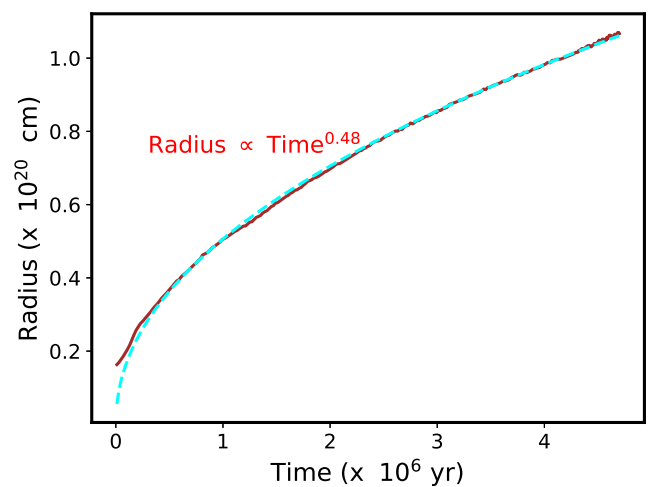}
\caption{The evolution of the outer shell radius over time (shown in brown) from the simulation described herein. Overplotted is a cyan curve with Radius $\propto $ Time$^{0.48}$. The~good fit clearly shows that the radius is not consistent with that predicted by the \citet{weaveretal77} model, but~is consistent with a momentum-conserving bubble. 
\label{fig:rad}}
\end{figure}

The calculation of the bubble radius at various epochs for~the simulation used in this paper (shown in Figure~\ref{fig:hydro}) shows that it is consistently less than the pressure-driven value proposed by \citep{weaveretal77} throughout most of the evolution.  In~Figure~\ref{fig:rad}, we plot the evolution of the bubble radius. Due to the various instabilities, an~average radius is difficult to obtain. The~radius shown here is  the radius of the bubble $R_b$ at $\theta=0^{\circ}$. In~the post-main sequence phases, the bubble does not grow significantly;~thus, the radius only increases slightly. As the wind parameters change between different stages, the~constant of proportionality in Equation~(\ref{eq:ec}) (the quantity within the square brackets preceding the time) will also change between phases. Conversely, these phases last for a short amount of time compared to the MS phase.   Throughout most of the evolution, we find that the radius of the bubble increases with time as $R \propto t^{0.48}$, keeping the above caveats in mind. The expansion rate  reaches the value of 0.48 by around 500,000 years. Prior to this, the rate is higher, as~would be expected for a pressure-driven bubble. The~500,000-year timescale reflects the time it takes for the clumps and filaments to form, grow, and cool, such that the expansion slows down from the pressure-driven value to the momentum-driven value. For~confirmation, we also checked the bubble radius  $R_b$ where $\theta=90^{\circ}$. Here, it is more difficult to measure due to a protrusion that exists along the axis, but~using an average radius, we obtain a similar value of the expansion rate, 0.5. The radius evolution shows that the application of a pressure-driven solution for the expansion rate of the bubble throughout its evolution, as~is generally done, is not appropriate for wind-blown bubbles around massive stars. The~bubble is more consistent with expansion in the momentum-conserving phase over most of its evolution.

\vspace{-4pt}

It is interesting to note that, as early as 1982, \citet{tc82} had pointed out that the observational properties of some ring nebulae around W-R stars agreed better with the momentum-conserving solution rather than the energy-conserving solution proposed by \mbox{\citet{weaveretal77}}.

\section{Energy Loss and Bubble Evolution}\label{sec4}

The expansion of the wind-blown nebula in the \citet{weaveretal77} description is due to the high pressure within the hot shocked wind region, which pushes out on the thin dense shell of swept-up material, causing the shell to expand outwards. The~pressure within the shocked wind region decreases with time. In~this model, the shocked wind region is adiabatic and unable to cool radiatively.
 The~pressure throughout the hot shocked wind region is uniform and high, while the density in the interior of the bubble is low, leading to  a high temperature. The~slower expansion derived in Section~\ref{sec:rad} suggests the existence~of various processes that could use up the energy required for the expansion of the bubble, thereby reducing the pressure within the hot shocked wind region.  We explore various factors that could decrease the available energy reservoir, causing the pressure within the shocked wind region to be lower than in the pressure-driven case. 

\begin{itemize}
\item{Instabilities:} The 2D simulations show the presence of several hydrodynamic and ionization front instabilities. Instabilities are an inherent feature of  multi-dimensional simulations. Paper 1 identified various instabilities found in different~regions. 
\begin{itemize}
\item The ionization front is found to be unstable to D-type ionization front instabilities.
These are mainly prevalent in the early evolution of the bubble (see Figure~\ref{fig:hydro}).
\item \textls[-20]{Finger-like projections, due to various instabilities, combined with photo-evaporative} absorption, are seen at the inner edge of the dense shell during the evolution. These are seen throughout the evolution.  
\item  The interface of the hot shocked wind region with the ionized HII region is susceptible to Rayleigh--Taylor and Kelvin--Helmholtz instabilities (Figure \ref{fig:hydro}). These instabilities grow throughout the evolution. The~hot shocked wind region itself does not maintain a spherical geometry. Mixing of the cooler HII region material (at $\approx$10$^4$ K) with the hotter bubble material can lead to cooling and an overall reduction in temperature in the hot bubble.
\item Hydrodynamic simulations \citep{dwarkadas07} have shown that the RSG wind shell is unstable to Rayleigh--Taylor instabilities. 
The~W-R wind expanding within the RSG wind is also unstable to the Rayleigh--Taylor (R-T) instability \citep{dwarkadas07}, 
with~the R-T fingers pointing inwards. The~high momentum W-R wind breaks apart the unstable RSG wind, leading to the formation of clumps and filaments that are mixed into the hot shocked wind region.
\end{itemize}
The various instabilities can lead to a high rate of cooling at the interfaces, which will be absent in spherically symmetric simulations. Mixing can occur between the dense ionized region and the shocked wind region, adding mass and cooler material to the hot shocked wind.  The~instabilities themselves are a function of grid resolution and~the hydrodynamic methods used to carry out the simulation; the number, size, and growth may vary between 2D and 3D simulations. A comparison between 2D and 3D instabilities in circumstellar shells around massive stars, without~photoionization, was carried out by \citep{vmk12}. Highly resolved 3D simulations are required to properly account for the effect of various instabilities. We plan to carry these out in~the future.

\item{Vorticity within the bubble:}  A wind termination shock separates the freely expanding wind from the shocked wind. The~position and~shape of the shock changes with time (Figure~\ref{fig:hydro}). The~shape of the shock front responds to inhomogeneities in the flow, the~presence of clumps and other density perturbations, and~hydrodynamical instabilities. This is also seen in simulations of planetary nebulae~\cite{db98a}, which are wind bubbles around lower-mass stars. The~change in the position and shape introduces vorticity within the bubble. The~vorticity deposition is carried out with the flow and~results in the formation of vortices in the shocked wind region. Figure~\ref{fig:vort} shows the velocity vectors (blue) imprinted over the density contours (red) for the simulation in Figure~\ref{fig:hydro}. The~formation of vortices in the shocked wind region is clearly visible. The~vortices are long-lasting, and~tend to cluster and sometimes merge together, forming larger vortices. This could be a function of the 2D nature of the simulations. In~2D, the energy cascades to longer wavelengths, unlike 3D simulations, where the energy is expected to cascade to smaller~scales.

\begin{figure}[H]
\hspace{-7pt}\includegraphics[width=0.98\columnwidth]{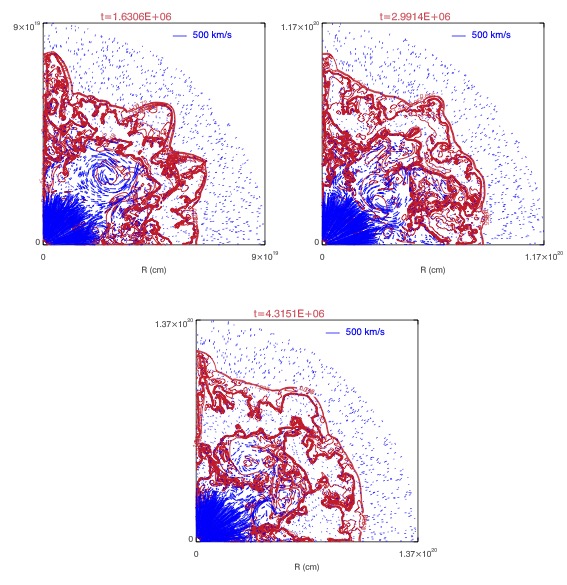}
\caption{Snapshots~ 
 showing the velocity vectors (in blue) plotted over the density contours (in dark red). The~formation of vortices in the hot shocked wind region is clearly seen. Time increases
 from left to right and is~listed in years at the top of each panel. The~vector in the top right corner of each panel denotes a velocity of 500 km s$^{-1}$. The~panels all correspond to the MS phase; the last one is at the transition between MS and RSG~phases.
\label{fig:vort}}
\end{figure}

The evolution of vorticity is shown mainly in the MS phase. In~the RSG phase, the~low velocity RSG wind does not travel far into the shocked bubble, while the high density of the wind leads to a new pressure equilibrium. In~the W-R phase, the~various instabilities, and~the transport of the RSG and W-R material into the shocked wind, make visualization of the velocity vectors and the vortexes difficult. Dwarkadas
 \citep{dwarkadas08} carried out a calculation of the energy in non-radial motions in the nebula, and~concluded that 15--20\% of the energy in the W-R phase goes into turbulent motions, thus reducing the energy that can go into heating the diffuse~gas. 

Two-dimensional (2D) turbulence is known to have properties that differ from 3D turbulence. Specifically, 2D turbulence results
in an inverse cascade in energy, as~opposed to 3D turbulence, where energy cascades to the lowest scales \citep{sy97,dg00,delbendeetal04,es04,se04,bfk07}. 2D turbulence also results in a clustering of long-lived vortices, which is not seen in 3D.  These results, while well documented, are generally based on experiments conducted with 2D incompressible flows. Our
simulations are axially symmetric and~involve  compressible flows. Therefore, the applicability of the results is unclear. As the motion and change in shape of the wind termination shock are physically motivated, qualitatively we expect similar results in 3D to those obtained herein. Vorticity deposition at the inner shock would be expected, although~the extent of the turbulence may differ. The~size and structure of vortices will differ in the 3D calculations as compared to our 2D ones. Therefore, the amount of energy expended in turbulence in 3D may be different from 2D, but~there is no doubt that turbulence will deplete the energy in the shocked wind region and~thereby the pressure driving the bubble~expansion. 

\item{Mass loading:} \cite{km92b} showed that the shocked wind in an adiabatic bubble with a radiative shell can transition to a partially radiative bubble only if there is additional mass injection. A~partially radiative bubble is one where the cooling time of the gas in the hot shocked wind is larger than the crossing time, but~smaller than the age of the bubble. In~the MS phase, hydrodynamic and ionization front instabilities lead to the formation of clumps and fingers that are injected into the wind bubble. The~size of the bubble, and~hence its volume, is mainly set in the main-sequence phase, while most of the mass emitted by the star, which is~mixed in with the shocked wind region, arises in the RSG and W-R stages. The~post-MS  phases occupy only about 10\% of the lifetime of the star, adding a substantial amount of mass to the bubble without a significant increase in volume. The~breakup of the RSG material by the W-R wind results in the formation of clumps and filaments that are mixed in with the hot shocked wind material. This increases the mass without changing the volume appreciably, thus enhancing the density. In~our simulations, the average density in the hot shocked wind region is still not large enough to make the cooling time smaller than the bubble age. However, there are regions of very high density, such as clumps and filaments, where the cooling time becomes shorter than the age of the bubble, leading to local cooling of the shocked wind material.

  \end{itemize}

\section{Discussion}\label{sec5}

While~ 
 a single simulation has been presented, simulations of bubbles around different mass stars carried out by us show similar instabilities, although~the size and growth of the instabilities may vary.  The~conditions necessary for the growth of the instabilities are prevalent in most wind-blown bubbles, as~are the conditions for vorticity deposition. Other simulations in the literature \citep{lancasteretal21b,geenetal21} have found that bubbles in various environments can be efficiently cooling. Gupta et al.
 \citep{guptaetal18} found in their simulations that the radius of bubbles around compact young star clusters was smaller in three dimensions than in one dimension, which would imply similar characteristics. However, they did not discuss it further. \mbox{\citet{georgyetal13}} found that the expansion rate of wind bubbles around rotating massive stars could be slower than the \citet{weaveretal77} value, which they attributed to various factors, including cooling, metallicity, as~well as the bubble coming into pressure equilibrium with the surroundings. It seems reasonable to assume that efficient cooling of WBBs is a general feature. In~our simulations, the loss of energy may occur due to hydrodynamic and ionization front instabilities; due to mass loading of the hot bubble; and due to the creation  of vortices. Thus, there are several avenues that can lead to dissipation of the bubble energy and pressure. \citet{lancasteretal21b}  suggested that turbulent mixing at the interface is the reason for the efficient cooling. Cooling at the interface due to instabilities certainly plays a role in our simulations, and~a reduction in temperature is seen near the interface between the hot shocked wind and the surrounding HII region,
 compared to the temperature near the wind termination~shock.

\begin{figure}[H]
\includegraphics[width=0.98\columnwidth]{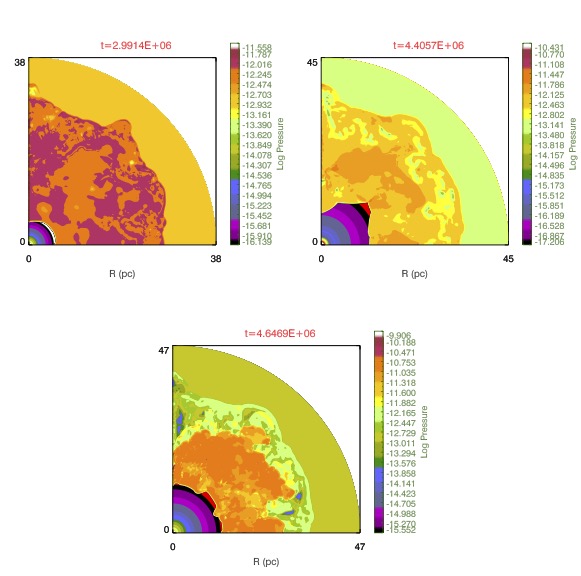}
\caption{Snapshots showing the pressure within the wind-blown bubble in various phases.  Note the variations in pressure within the shocked wind region, as~opposed to the uniform pressure expected in the \citet{weaveretal77} model. Time (in years) increases from left to right and~is listed at the top of each panel. The~first panel corresponds to the MS phase, the~second to the RSG phase, and~the third to the W-R~phase.
\label{fig:pre}}
\end{figure}

Figure~\ref{fig:pre} shows the pressure within the wind bubble throughout its evolution. Unlike the spherically symmetric case, the~pressure within the shocked wind region, as well as within the dense ionized HII region, is no longer uniform, but~is seen to vary spatially. The~pressure in the hot bubble reaches an equilibrium with the pressure in the HII region during the MS and RSG phases, but~is higher in the W-R phase. Pressure variations within the ionized region are clearly visible in the middle and rightmost panels and~coincide with the sites of instability formation in the region. Pressure variations within the shocked wind region are apparent throughout the evolution. In~the top
 left panel, depicting the MS phase, a~decrease in pressure is seen near the position of the vortices, as~well as in regions near fluid instabilities. Low pressure near the interface where instabilities are present, and~near the vortices, is also visible in the top right panel. In~the bottom panel, depicting the W-R phase, the~pressure is lower near where the clumps and filaments are formed due to the W-R wind crushing the RSG wind (see Figure~\ref{fig:hydro}). The~lower pressure in various regions results in a reduction of the overall pressure, thereby decreasing the expansion of the~bubble. 

In the past, thermal conduction at the interface between the hot shocked wind and the surrounding region has been proposed as an avenue for reducing the temperature \citep{chu08}. We do not include thermal conduction in our simulations but~did not find its inclusion necessary to reproduce the X-ray temperatures and~spectra.

The fact that WBBs around massive stars are cooling efficiently, and~are better described by a momentum-driven solution rather than the pressure-driven solution, has strong implications in various areas. We discuss some of these~below.

\begin{itemize}
\item{\bf Dynamics and Kinematics of WBBs around massive stars:
} If WBBs are better described by a momentum-conserving solution, then the radius of the bubble is smaller than that expected from \citet{weaveretal77}. Conversely, given the bubble radius $R$ and the shell velocity $v$ in~the \citet{weaveretal77} solution, the time of expansion (the age of the bubble), assuming a constant velocity, is given by
\begin{equation}
{t_{age}}_{PD}=\frac{3}{5}\frac{R}{v}
\end{equation}

\noindent
while in the momentum-conserving (efficiently cooling) regime, it is given by
\begin{equation}
{t_{age}}_{MC}=\frac{1}{2}\frac{R}{v}
\end{equation}

Thus, the momentum-conserving bubble will have a smaller age for a given radius and velocity. This can be used to resolve various discrepancies pointed out in the~literature.

\citet{nazeetal01} compared the expansion dynamics of interstellar bubbles in N180B and N11B with the \citet{weaveretal77} PD solution, and~found that it leads to wind luminosities that are at least an order of magnitude lower than~expected. 

The bubble blown by the O3-O4 star MGSD 214 in N180B has a radius of 11 pc and an expansion velocity 20 km s$^{-1}$, giving a dynamical timescale of 3.3 $\times 10^5$ yr. With~an rms density of $\sim$9.5 cm$^{-3}$, using the \citet{weaveretal77} solution they calculated a~wind luminosity of $\sim$3 $\times 10^{36}$ erg s$^{-1}$ for MGSD 214, which is a factor of 10 lower than expected from an O3-O4~star. 

Using the momentum-conserving bubble solution, we obtain an age that is 1.2 times 
lower, $\sim 2.6 \times 10^{5}$ yr. Using the expression in Equation~(\ref{eq:ec}), and~assuming a wind velocity of 2000 km s$^{-1}$, we obtain a wind luminosity of 6 $\times 10^{37}$ erg s$^{-1}$, which is comparable to what is expected from an O3-O4 star.

\citet{nazeetal01} similarly considered the  bubble blown by PGMW 3204,
3209, and 3223~in N11B. The~star PGMW 3209, although~dominated by an O3 III star, is in a cluster of at least 5 other O stars. Using the derived
radius of 7 pc, and~an expansion velocity of 10 km s$^{-1}$, they found a 
dynamical timescale for this bubble to be 4.1 $\times 10^5$ yr using the \citet{weaveretal77} solution. With~the
rms density of 15 cm$^{-3}$ for N11B, they found  a wind luminosity of $\sim$2.5 $\times$ 10$^{35}$ erg s$^{-1}$ for PGMW 3209, which is almost 2 orders of magnitude lower than what is expected from an O3 III~star. 

The momentum-conserving solution, on the other hand, gives a lifetime of 3.42 $\times 10^5$ yr. Using Equation~(\ref{eq:ec}), and~a wind velocity of 3245 km s$^{-1}$ for an O3 III star \citep{pc98}, we find a bubble luminosity of 1.6 $\times 10^{37}$  erg s$^{-1}$, bringing it in line with the expectation for an O3 III star, and~close to two orders of magnitude  higher than the value obtained by \mbox{\citet{nazeetal01}}.

\item{\bf Expansion velocities of the bubbles:} 

The velocity of a bubble $v_{MC}$ in the momentum-conserving phase is given by:
\begin{equation}
V_{MC} = 0.5\left[\frac{3}{2\pi}\frac{\dot{M}\,v_w}{\rho}\right]^{1/4}\;t^{-0.5}  = 5.81 \times 10^8 \, {{ t}_{yr}}^{-0.5}\; {\rm cm} \; {\rm s}^{-1}
\label{eq:vec}
\end{equation}

\noindent
where $t_{yr}$ is the time in years, and~we use $\dot{M} = 10^{-6} \msun$ yr$^{-1}$, \mbox{$v_w = 1000$ km s$^{-1}$}, and a density in the surrounding medium $\rho = 1.67 \times 10^{-24}$ cm$^{-3}$. Initially,
 the bubble will be energy-conserving, but~due to efficient cooling, it slowly transforms into a momentum-conserving bubble. The~time when the momentum-conserving phase is reached will vary for different mass stars. Equation~(\ref{eq:vec}) applies only after the momentum-conserving stage is reached.  In~our simulation, this happens at around half a million years. From~that time onward the velocity in our simulation (for the radius shown in Figure~\ref{fig:rad}) is found to decrease, as t$^{-0.51}$, which is consistent with the value for a momentum-conserving bubble.   Prior to this, the bubble will have a higher velocity, being in the pressure-driven stage.  Therefore, although~the velocity decreases at the rate given by  Equation~(\ref{eq:vec}) after around 500,000 years, its actual value is slightly higher than that calculated from the above equation. The~equation implies that in a million years the velocity should decrease to 5.8 km s$^{-1}$.  The~velocity and mass-loss rate of the stellar wind will change with time, so the above estimate is an approximation, but~it clearly shows that the velocities of wind bubbles will be low throughout most of the star's evolution. The~velocities of typical wind bubbles, both on the MS and in the W-R phase, are found to be generally less than 10 km s$^{-1}$ \citep{cappaetal03}, so these values are consistent. The~bubble may also approach pressure equilibrium with the surrounding medium. For~a higher ambient density, the~bubble velocity scales as ${\rho}^{-1/4}$, whereas it will increase with increasing mass-loss rate and wind velocity as ${\dot{M}}^{1/4}$ and ${v_w}^{1/4}$, respectively.

\citet{cgg03} contend that the low velocity implies weak shocks and a lack of compression of the material in the dense shell, which  could account for the lack of observed MS bubbles, as~it will make the bubble difficult to observe at optical wavelengths, especially if the surrounding medium is an HII region.  This is debatable, as~the slow shock will likely be a radiative  shock, as 
 expected in wind bubbles, with~a total compression ratio larger than that for a  strong shock. This will happen until the shock is close to reaching pressure equilibrium with the surrounding medium. Thus, a lack of compression cannot be the reason. Chu et al.
 \citep{cgg03}  further contend that as massive stars evolve, they will `lose ionizing power', and~the bubbles and the ambient medium will recombine and cool, making them more detectable. There are difficulties with this argument when comparing MS and W-R bubbles. W-R stars have more ionizing photons than MS stars, so the argument that stars lose ionizing power as they age would not hold, in~fact they become more powerful ionizing~sources. 

We suggest here that the reason that W-R nebulae, formed when stars age, are more easily detectable than MS nebulae in the optical, is simply because their optical
 luminosity is higher. For~one, the~dense shell has swept up a larger amount of mass ($\propto R^3$) by the W-R stage. It has also expanded outwards, and~its radius  has increased. The~shell  volume $\propto R^2 \Delta R$, where $\Delta R$, the~thickness of the dense shell, increases almost proportionately to the radius in self-similar evolution. Therefore, the density remains the same or decreases slightly. Due to the larger volume, the~H$\alpha$ emission resulting from recombination in the shell will be higher. Secondly, the~W-R wind carries both its mass and the mass of the prior RSG wind, mixing it into the hot shocked MS wind. As~pointed out in Paper 1, and in earlier sections, the~W-R wind collides with the dense RSG wind, breaking it up, and~forming high density clumps and filaments in the hot shocked wind region.  The~densest clumps are too dense to emit in X-rays but~have the right temperature to emit in the optical (Figure \ref{fig:temp}). The~H$\alpha$ luminosity arising from Case B recombination is, similar to the X-ray luminosity (Section~\ref{sec:Xray}), a~function of the square of the plasma density. Therefore, the~densest clumps can emit in H$\alpha$, and~the H$\alpha$ luminosity in the W-R phase will be significantly higher than in the MS phase, contributing to the detectability at optical wavelengths. We therefore suggest that it is the higher optical luminosity of bubbles in the W-R phase, and~not the shock compression, which makes W-R bubbles more detectable compared to MS~ones.

\end{itemize}

\begin{figure}[H]
\includegraphics[width=0.98\columnwidth]{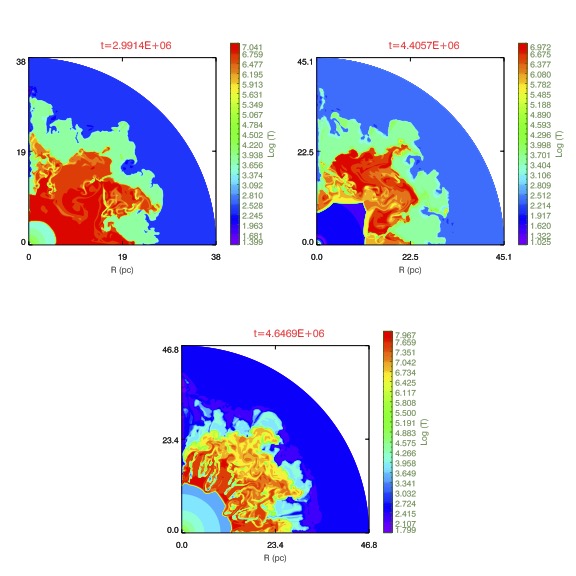}
\caption{Snapshots showing the temperature of the wind-blown bubble in various phases.  Note the variations in temperature in the hot shocked wind, with~low temperatures particularly at the sites of instabilities, and mixing of cool ionized material with the hot interior. Time
 (in years) increases from left to right and~is listed at the top of each panel. The~first panel corresponds to the MS phase, the~second to the RSG phase, and~the third to the W-R~phase.
\label{fig:temp}}
\end{figure}

\section{X-ray Emission from Wind-Blown Bubbles}
\label{sec:Xray}

As mentioned above, both the X-ray temperatures and the X-ray luminosity observed in bubbles, especially W-R bubbles, are lower than expected. In~Figure~\ref{fig:temp} we show the temperature calculated from our simulations during various stages of the evolution. The time in years is given at the top. The~temperature shown is calculated by assuming a mean molecular weight $\mu$ of 0.6, appropriate for a fully ionized region. In~principle, it should be calculated assuming a mean molecular weight appropriate for the abundances, which could vary with time. However, this approximation will suffice, as~the variation in $\mu$ will be less than a factor of 2, and 
 it is the spatial variation in temperature that we wish to~emphasize.

The maximum temperature in the MS phase is found to exceed 10$^7$ K, although~there are variations, primarily near the interface of the hot shocked wind with the HII region, where the growth of instabilities as well as the mixing of cooler HII region material with the bubble interior is expected to occur. In~the RSG phase, as~the medium begins to recombine, the~temperature decreases, both in the ionized region and in the hot shocked wind material.  In~the W-R phase, the~ stellar surface temperature increases, and~the hot shocked wind temperature rises correspondingly, reaching a maximum temperature close to 10$^8$~K. However, the~temperature varies throughout the plasma in the shocked wind region, with~the temperature in the outer parts, near~the interface with the ionized region, being an order of magnitude lower than that close to the reverse shock. This is presumably due to mixing of material at the interface as a result of various instabilities, plus the fact that at this epoch, the W-R wind has not yet penetrated the outer parts of the shocked wind region, which mainly consists of MS material. As~the W-R wind initially expands at a high velocity (compared to the prior RSG wind), it quickly encounters the piled-up RSG wind material. The~W-R wind collides with the RSG material, breaking it apart due to the higher momentum of the W-R wind. This results in the formation of filaments and clumps within the hot bubble. These have the highest density, and~the lowest temperature, with~some having a temperature as low as 10$^{3}$ K, as~can be seen in Figure~\ref{fig:temp} in the bottom panel.

If one considers the average temperature in the entire shocked wind region in the W-R phase, it will be $\ge$10$^7$ K, even with efficient cooling. This by itself cannot explain X-ray temperatures near 10$^6$ K seen in the observations. We explore what sets the low temperature and lack of observed diffuse X-ray emission in most X-ray~observations.

The X-ray flux is proportional to $n_e^2\,T^{1/2}$, where $n_e$ is the density and $T$ is the temperature.  Compared to the spherically symmetric model assumed by \citet{weaveretal77}, the volume of the shocked wind region is reduced due to the smaller radius of the bubble overall, as~well as the aspherical geometry of the shocked wind region. The~density of the plasma will proportionately increase, but~as mentioned above, the highest density clumps have too low a temperature to contribute to the X-ray emission. The~average density of the remaining high-temperature material in the hot shocked wind region that can contribute to the X-ray emission is, therefore, lower than expected.  Given this density and the smaller volume, the~overall emissivity is reduced compared to the model~expectations. 

 During the MS phase, the mass loss rate is low, and~so a small amount of mass is lost. As~is clear from Figure~\ref{fig:temp}, in~the shocked wind region, there exist some high density clumps with temperatures below 10$^6$ K that do not contribute to the X-ray flux. While the average temperature is $\geq$10$^7$ K, the~density of the remaining material is lower than expected from the model, and~the overall luminosity of the hot shocked wind is generally too low to produce detectable~emission. 
 
In the W-R phase, we can divide the plasma into three components: (1) The highest density clumps (those in dark green or light blue in Figure~\ref{fig:temp}, bottom panel). These  have temperatures below 10$^6$ K, and~as low as 10$^3$ K. They do not emit in X-rays and~do not contribute to the X-ray flux. (2)  High density plasma in filaments and clumps, with~temperatures just above 10$^6$ K, especially material that appears light-greenish to yellowish in color in Figure~\ref{fig:temp}, bottom panel. These give rise to the majority of the X-ray flux.  These are dense regions in the plasma (compare to Figure~\ref{fig:hydro}, bottom right), with~densities much higher than the average plasma density in the hot shocked wind. They arise mainly from instabilities during the various phases, forming filaments that break from the interface, and are mixed in with the plasma; or are~due to the W-R wind mixing with the RSG wind. The~flux scales as the square of the density, making the filaments and clumps the largest contributors to the X-ray emission, despite their low volume filling factor.  In~the MS phase, there are few such dense filaments and clumps; thus, the X-ray emission in the MS phase is not similarly enhanced.  (3) The remaining lower density plasma in the shocked wind region. Much of the mass from the star is lost in the post-MS phases, and~is mixed in with the hot shocked wind region. A~substantial amount goes into clumps and filaments, with~the rest going to increase the average density of the hot shocked wind region compared to the MS phase. This component will give rise to high temperature ($\approx$10$^7$ K), low luminosity X-ray emission from the nebula. The~contribution from this component is initially small, but~as pointed out below, may increase with~time.   

Since the majority of the flux arises from filaments whose temperatures are just above 10$^6$~K (Figure \ref{fig:temp}), it is not surprising that the observed emission in W-R nebulae  peaks slightly above 10$^6$ K. The~rest of the hot shocked region, with~a temperature $\geq$10$^7$ K, can make a small higher temperature contribution to the observed luminosity, but~it is not a significant fraction due to the lower density compared to the~clumps.

\begin{figure}[H]
\includegraphics[width=0.98\columnwidth]{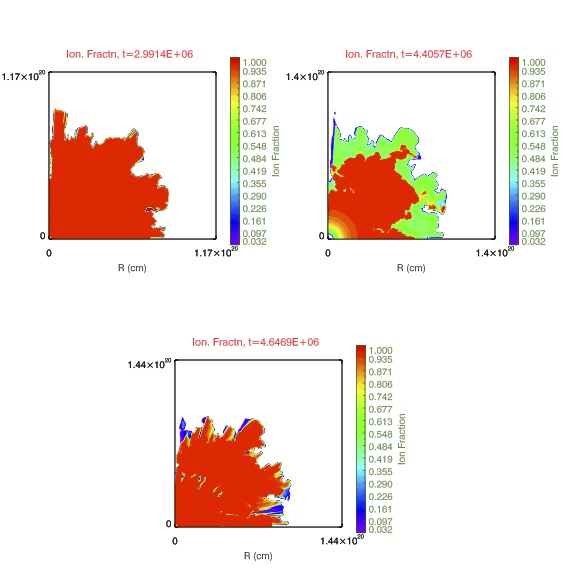}
\caption{Panels  
  showing the ionization fraction within the wind-blown bubble over time. Time  (in years) increases from left
  to right, and~is listed at the top of each panel.  The~first panel corresponds to the MS phase, the~second to the RSG phase, and~the third to the W-R phase. The~ionization fraction decreases in the RSG phase, then steadily increases in the W-R phase.
\label{fig:ion}}
\end{figure}

Due to their high density, many of the densest clumps are not fully ionized (Figure~\ref{fig:ion}). They tend to shield plasma directly behind them from the ionizing effect of the stellar photons. The~low ionization plasma can absorb the X-ray emission arising from a region interior to the clumps, which is the hot shocked material close to the wind termination shock. This will further reduce the observed X-ray emission arising from this region while increasing the column~density.  

\vspace{-3pt}

The lack of observed diffuse X-ray emission in the W-R phase compared to that expected from the \citet{weaveretal77} model is due to factors such as the lower emitting volume compared to the model, as~pointed out above, as~well as the ionization fraction of the gas. The~ionization fraction in the nebula is depicted at various times during the evolution in Figure~\ref{fig:ion}. Note that our simulations only include the ionization of H. In the main-sequence phase, we see that the entire bubble is almost fully ionized. In~the RSG phase, the~temperature of the star is low and unable to ionize the entire nebula, and~the ionization in the outer HII region begins to decrease. The figure shows that the average  ionization fraction is around 0.5, although~in some regions it is lower than 0.2.  By~the end of the RSG phase, the average ionization fraction in the ionized HII region is found to be lower than 0.3.  This means that at the beginning of the W-R phase, the~outer partially ionized and neutral plasma of the nebula is able to absorb a large fraction of the X-ray emission. As~time proceeds, the W-R star with its high surface temperature begins to ionize the nebula, and~the column density decreases over time. However, this takes time, and~there is material in a low state of ionization in the outer parts that is able to absorb some amount of the X-rays.  Even at very late times, close to the end of the star's life, as~seen in the bottom panel in Figure~\ref{fig:ion}, there is still some material with an ionization fraction of $\leq$0.25, adding to the absorption. Thus, absorption of the X-ray emission contributes to lowering the observed X-ray~luminosity.

\begin{figure}[H]
\includegraphics[width=0.98\columnwidth]{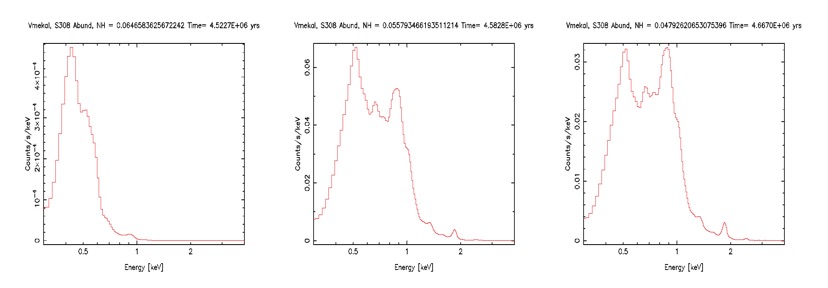}
\caption{  Evolution of the X-ray spectra within the W-R phase of the bubble.  Time increases from left
  to right. The~absorption column due to the nebular material decreases with time. Adapted from Figure~2 
 in citep{dr13}.
\label{fig:wrspec}}
\end{figure}

The decreasing column density in the W-R phase can be seen in our simulated X-ray spectra in the W-R phase, as shown in Figure~\ref{fig:wrspec}, which is adapted from Figure~2
 in~\citep{dr13}.  The~ISIS package (\citep{houck2000}) is used to calculate the X-ray spectra from the simulations.  The~output data from the simulation are read into ISIS. The~spectra are calculated for every grid cell, taking the absorption outside that cell into account, and~added together.  The~calculations assume a point source at 1.5~kpc distance, integrated over 50,000~s, and~then convolved with the response of the Chandra ACIS-S  instrument. The~VMEKAL model, an~ionization equilibrium model that takes both thermal bremsstrahlung and line emission into account, is used to model the spectrum. It is possible that in newly shocked regions, the plasma may not be in ionization equilibrium, and~non-equilibrium ionization models need to be calculated. Given the age of the bubble, the~size of such a region is expected to be quite small, and~therefore for illustrative purposes we can neglect it. The~absorption column for any grid cell is calculated by summing over all cells beyond that cell in radius. The~column density from all cells is then added to obtain the total absorption column within the nebula. To this value is added a foreground absorption of 2 $\times$ 10$^{20}$ cm$^{-2}$ in the direction of the source. The~total N$_H$ is given in units of 10$^{22}$ cm$^{-2}$. Line broadening is based on the underlying fluid velocity.  Solar abundances (\cite{ag89}) are used for the MS and RSG stages. Abundances in the W-R phase are from~\cite{chuetal03} for the W-R bubble S308.  The~counts s$^{-1}$ keV$^{-1}$ in the main sequence phase  amount to a few times 10$^{-4}$ (not shown), while those in the W-R phase (shown in Figure~\ref{fig:wrspec}) increase by almost two orders of magnitude, to a few times 10$^{-2}$ counts s$^{-1}$ keV$^{-1}$. The~spectral shape is found to be consistent with observed spectra of S308 and NGC 6888~\cite{chuetal03, zp11}. The~title at the top of each panel lists the X-ray model used (VMEKAL), the~abundance used (S308), the~column density (NH), and the time in years. The~decrease in the absorption column over the W-R phase is~evident.

As the evolution proceeds, the~W-R wind tends to fill the entire hot shocked wind region, and~the temperature of the shocked wind region increases. This, combined with the decreasing column density, can lead to the emergence of a higher temperature component in the spectrum, as seen in our simulated X-ray spectra (Figure~\ref{fig:wrspec}). This component initially has a much lower luminosity, although~its intensity increases with time. Higher temperature components which contribute a small fraction of the total X-ray luminosity have been observed in NGC 2359 and NGC 3199.

Therefore, we  contend that it is the tendency of the majority of the X-ray emission to originate from high density clumps and filaments at temperatures slightly above 10$^6$ K which leads to  the low observed X-ray temperature in the W-R phase. Emission from the rest of the nebula may make a small contribution, resulting in a higher temperature component that constitutes a small fraction of the luminosity, which may increase with~time.

\section{Conclusions}\label{sec7}

In this paper, we further analyzed the multi-dimensional ionization-gasdynamic simulations presented in Paper 1. Our results clearly show  that various energy dissipation processes are at work in wind bubbles around massive stars, leading to efficient cooling. The~various energy leakage sources include cooling at the interfaces due to various instabilities and  mixing of hot and cool material, coupled with instabilities at the ionization front, and~the formation of vortices within the interior of the hot shocked wind. The~pressure within the bubble varies spatially, unlike in the \citet{weaveretal77} model. The~temperature is considerably reduced at the site of fluid instabilities and mixing of cool material with the hot medium; temperature reduction is also observed~near the formation of clumps and filaments, which is consistent with the observation that those are the sites associated with energy loss.  Strong cooling tends to reduce the radius of the bubble, as~calculated from our multi-dimensional simulations. The~evolution of the wind bubble is more comparable to the momentum-conserving case than to the energy-conserving scenario proposed by \citet{weaveretal77}.

Efficiently cooled bubbles have been discussed in the literature in various \linebreak contexts~\mbox{\citep{ssw75, km92a, km92b, st13, mackeyetal15, lancasteretal21a}}. Our results are consistent with the results derived for bubbles formed around a cluster of stars \citep{lancasteretal21b}, which do not include the effect of stellar ionizing photons. \citet{geenetal21} also found that wind bubbles evolve to an efficiently cooled system.  It is clear that realistic wind-blown bubbles around massive stars, whether single or in a cluster, are generally
 not energy conserving, and~their evolution is not well described by the \citet{weaveretal77} solution. This has far reaching implications. We show that the momentum-conserving solution can better explain the dynamics and kinematics of wind bubbles around massive stars, resolving discrepancies that have been discussed in the literature. It can better recover the mechanical luminosity of  observed bubbles, as~well as explain the low velocities of bubbles around stars on the main sequence.

The \citet{weaveretal77} solution has  been used to model wind-blown bubbles in many scenarios. It is used to describe the dynamics of superbubbles \citep{mm88} and model superbubbles under various conditions \citep{guptaetal16}, and~has further been employed to model cosmic-ray acceleration within superbubbles \citep{parizotetal04, vieuetal22}. It has also been used to study particle acceleration in star clusters \citep{morlinoetal21}. We suggest that the calculations in many of these scenarios could change, often significantly, if~a momentum-conserving solution is used to describe the bubble~evolution.

In this work, we did not take the rotation of the star into account. Rotation could lead to aspherical mass loss from the star, consequently influencing the shape of the bubble \citep{chitaetal08, vanmarleetal08}. Slow rotation, with~a  velocity much less than the critical velocity, would probably not have a significant impact. Rotation at velocities comparable to the critical velocity can have a significant impact. Maeder and Desjacques \citep{md01} and Dwarkadas and Owocki \citep{do02} 
have shown how stars rotating near critical velocity could result in the formation of bipolar bubbles. Such bubbles have not been seen around W-R stars, suggesting that they are not very fast rotators.  On~the other hand, bipolar nebulae are regularly seen around LBV stars \citep{weis11, gm12, gvaramadzeetal15, wb20}, perhaps suggestive of a high rotation velocity. A~bipolar wind-blown nebula was also seen around SN 1987A. The~latter  has been attributed to a higher density at the equator compared to the poles \citep{bl93}, but~rotation could play a factor.   Shaping of nebulae around rotating massive stars,  without~taking stellar photoionization into account, has also been discussed by \citep{georgyetal13}.

Although the bubble is efficiently cooling and its radius is smaller than in the \linebreak \mbox{\citet{weaveretal77}} case, the~temperature within the shocked wind region, although~varying by more than an order of magnitude, is not low enough in our calculations to account for the low observed temperatures. We argue that most of the X-ray emission arises from small regions consisting of dense clumps and filaments, which result from instabilities at the interface, as~well as the collision of the W-R wind with the piled-up RSG material from the previous epoch, and~the subsequent destruction of the RSG wind material. These clumps have densities up to two orders of magnitude higher than the surrounding hot shocked material, and~temperatures just above 10$^6$ K. Even denser clumps with temperatures below 10$^6$ do not contribute. The~resulting X-ray emission mainly reflects these  low temperature clumps. The~remainder of the shocked wind region can make a higher temperature contribution, although it will be small compared to that from the lower temperature clumps and filaments. Absorption of the X-ray emission in the beginning and early stages of the W-R phase can contribute to the low observed X-ray~luminosity.

\citet{ta18} \hl{} studied the X-ray temperature of hot gas in diffuse nebulae. They suggest that in all cases, turbulent mixing layers transfer energy from the hot shocked stellar wind to the photoionized gas. While this is also true in our simulations, we have shown that it is only part of the story. Instabilities are also present at the interface between the photoionized gas and dense swept-up shell, and~the shell itself is unstable to D-type ionization front instabilities. 
Cooling at all of these interfaces results in a loss of energy. Energy losses may also occur due to turbulence and mass-loading as described herein. As~a result of these, the~nebula begins to expand at the slower momentum-conserving rate, leading to a reduction in nebular size compared to the energy-conserving solution. This result was not previously noted in calculations of nebulae around W-R stars. The~smaller size may result in a  slightly larger plasma density, but~not all of this dense plasma contributes to the X-ray emission. Some of it is too dense to emit at X-ray temperatures. Furthermore, while \citep{ta18} attribute the majority of the X-ray emission to these turbulent mixing layers, we find that at least part of the X-ray emission in W-R nebulae arises from dense clumps and filaments formed by the collision of the W-R wind with the wind material from the previous epoch, which in~this case is the RSG wind. These clumps are mixed in with the hot shocked wind region. \citep{ta18} suggest that a second temperature component could be present due to the shocked fast wind material, which we also find in our simulations. 

Wind-blown bubbles are also found around low mass stars. These are the planetary nebulae (PNe), formed by the interaction of a fast central wind with the slower AGB wind from a previous epoch. Therefore, it is inevitable that comparisons will be made between PNe and the bubbles around massive stars. However, we caution that there is a fundamental difference, which is that PNe are much shorter-lived structures, lasting for perhaps \mbox{10,000--20,000 years}. This results in two major differences compared to WBBs around massive stars: (1)~There is generally not enough time for the plasma to cool radiatively in most cases, even though hydrodynamic instabilities may arise at the interface. (2)~The shocked plasma in PNe will not have time to reach ionization equilibrium, as~shown in \mbox{\citet{ssw08}}. Therefore, non-equilibrium ionization processes need to be used to accurately calculate the X-ray emission from the hot bubble,~specifically the X-ray spectra, making  the calculation more difficult and time-consuming. However, many calculations of the X-ray spectra from PNe have used ionization equilibrium processes~\mbox{\citep{ssw08, ta16}}, casting some doubt on the results. X-ray emission from PNe has been attributed to many different factors: (1)~Stellar winds in the early phases~\cite{sk03}. (2) The wind from the central star, or~collimated fast winds from the companion star \citep{ams08}. (3) Thermal conduction at the interface. In~our simulations, we did not find it necessary to include thermal conduction in order to explain the spectra of WBBs around massive stars. \citet{ta16} found that simulations both with and without thermal conduction can reproduce the X-ray temperatures and luminosities of PNe. As~mentioned above, \citet{ta18} suggest that turbulent mixing layers are responsible. \citet{raananetal09} interpreted a  radiative recombination continuum feature in the X-ray spectrum of a PN as evidence of charge exchange from the hot shocked wind to the colder nebular shell. Our simulations are consistent with turbulent mixing and energy transfer from the hot shocked wind. Due to the differences between WBBs around massive stars and PNe cited above, we will refrain from making further direct comparison to PNe at this stage. In~future, we plan to apply the techniques outlined herein to simulate PNe and~compute the X-ray emission using non-equilibrium ionization conditions.


\vspace{6pt}


\funding{V.V.D.'s  research is funded by the NSF grant 1911061, and previously by grants TM9-0001X and TM5-16001X provided by NASA through the Chandra X-ray Observatory center. The center is operated by SAO under NASA contract NAS8-03060.}

\dataavailability{The simulations described were previously published in Paper 1. The data described in the article were obtained through post-processing and analyzing the simulations, and are well-detailed. Anyone interested in accessing the actual simulations can request them from the author upon reasonable~\mbox{request}.} 

\acknowledgments{We thank the referees for a careful reading of the paper, and for their comments and suggestions, which greatly helped to improve the manuscript. V.V.D. is deeply grateful to Duane Rosenberg for generously providing the code and the tremendous assistance in updating and running the simulations used in this work.}

\conflictsofinterest{The author declares no conflict of interest.}


\abbreviations{Abbreviations}{
The following abbreviations are used in this manuscript:\\

\noindent 
\begin{tabular}{@{}ll}
WBB & wind-blown bubble\\
MS & main sequence\\
RSG & red supergiant\\
W-R & Wolf--Rayet \\
R-T & Rayleigh-Taylor \\ 
\end{tabular}
}

\appendixtitles{no} 
\begin{adjustwidth}{-\extralength}{0cm}

\reftitle{References}



\PublishersNote{}
\end{adjustwidth}
\end{document}